\documentclass[12pt]{iopart}

\newcommand{\sqrtsnn}{\sqrt{s_{_{\rm NN}}}}
\usepackage[pdftex]{graphicx}
\usepackage{amssymb}
\begin{document}

\title[Probing nPDFs and parton energy loss through $\gamma +Q$ production]{Probing nuclear parton densities and 
parton energy loss   processes through photon + heavy-quark jet production in p-A and A-A collisions}

\author{Tzvetalina Stavreva $^1$, Fran\c cois Arleo$^2$ and Ingo Schienbein$^1$}

\address{$^1$ Laboratoire de Physique Subatomique et de Cosmologie, UJF, CNRS/IN2P3,
\\
INPG, 53 avenue des Martyrs, 38026 Grenoble, France} 

\address{$^2$ Laboratoire d'Annecy-le-Vieux de Physique Th\'eorique (LAPTH), UMR5108,\\
 Universit\'e de Savoie, CNRS, BP 110, 74941 Annecy-le-Vieux cedex, France\\}
\eads{\mailto{stavreva@lpsc.in2p3.fr}, \mailto{arleo@lapp.in2p3.fr}, \mailto{schien@lpsc.in2p3.fr}}
\vspace{-0.2cm}
\begin{abstract}
We present a detailed phenomenological study of the associated production of a prompt photon and a heavy-quark jet (charm or bottom) in proton-nucleus (p-A) and nucleus-nucleus (A-A) collisions. The dominant contribution to the cross-section comes from the gluon--heavy-quark (gQ) initiated subprocess, making this process very sensitive to the gluon and the heavy quark nuclear parton densities. We show that the future p-A data to be collected at the LHC should allow one to disentangle the various nPDF sets currently available. In heavy-ion collisions, the photon transverse momentum can be used to gauge the initial energy of the massive parton which is expected to propagate through the dense QCD medium produced in those collisions.  The two-particle final state provides a range of observables (jet asymmetry, photon-jet pair momentum, among others), through the use of which a better understanding of parton energy loss processes in the massive quark sector can be achieved, as shown by the present phenomenological analysis carried out in Pb-Pb collisions at the LHC.
\end{abstract}

\vspace{-1cm}
\section{Introduction}
\vspace{-0.2cm}
The production of a prompt photon in association with a heavy-quark jet provides us with the opportunity to study the structure of the proton and the nucleus as well as the mechanisms of heavy quark energy loss. The information obtained depends on the collision type.
\vspace{-0.2cm}
\begin{itemize}
\item For  $p-\bar p$  collisions (at the Tevatron) it was shown in {\cite{Stavreva:2009vi}} that this process is sensitive to the charm/bottom PDF, and therefore can provide information and constraints on the presence of intrinsic charm/bottom (IC/IB) in the proton {\cite{Abazov:2009de}}. 
\item In  $p-A$ collisions (at RHIC and the LHC) $\gamma + Q$ production can be used to constrain the gluon nuclear PDF (nPDF), {\cite{Stavreva:2010mw}}, which presently carries a large error to it, as will be shown in more detail in Section \ref{sec:pA}. One should underline that knowing the precise nPDFs is necessary for obtaining reliable predictions in   $A-A$  collisions. 
\item In  $A-A$  collisions the study of prompt photons and heavy quarks provides an ideal tool for investigating the energy lost by heavy partons in the hot medium (Section \ref{sec:AA}).  As an electromagnetic probe the photon is expected to traverse the medium unaffected and thus gauge the quenching of the energy of the heavy-quark jet.  Furthermore, the comparison between $\gamma + c$ and $\gamma + b$ production provides access to the mass hierarchy of parton energy loss.
\end{itemize}
\vspace{-0.8cm}
\section{Constraining the gluon nPDF through $\gamma+Q$ production} 
\label{sec:pA}
\vspace{-1.32cm}
%%%
\begin{figure*}[ht]
\begin{center}
\begin{picture}(480,140)(0,0)
 \put(15,-15){\mbox{\includegraphics[angle=0,scale=0.26,trim= 0cm 0cm 0cm 2cm,clip]{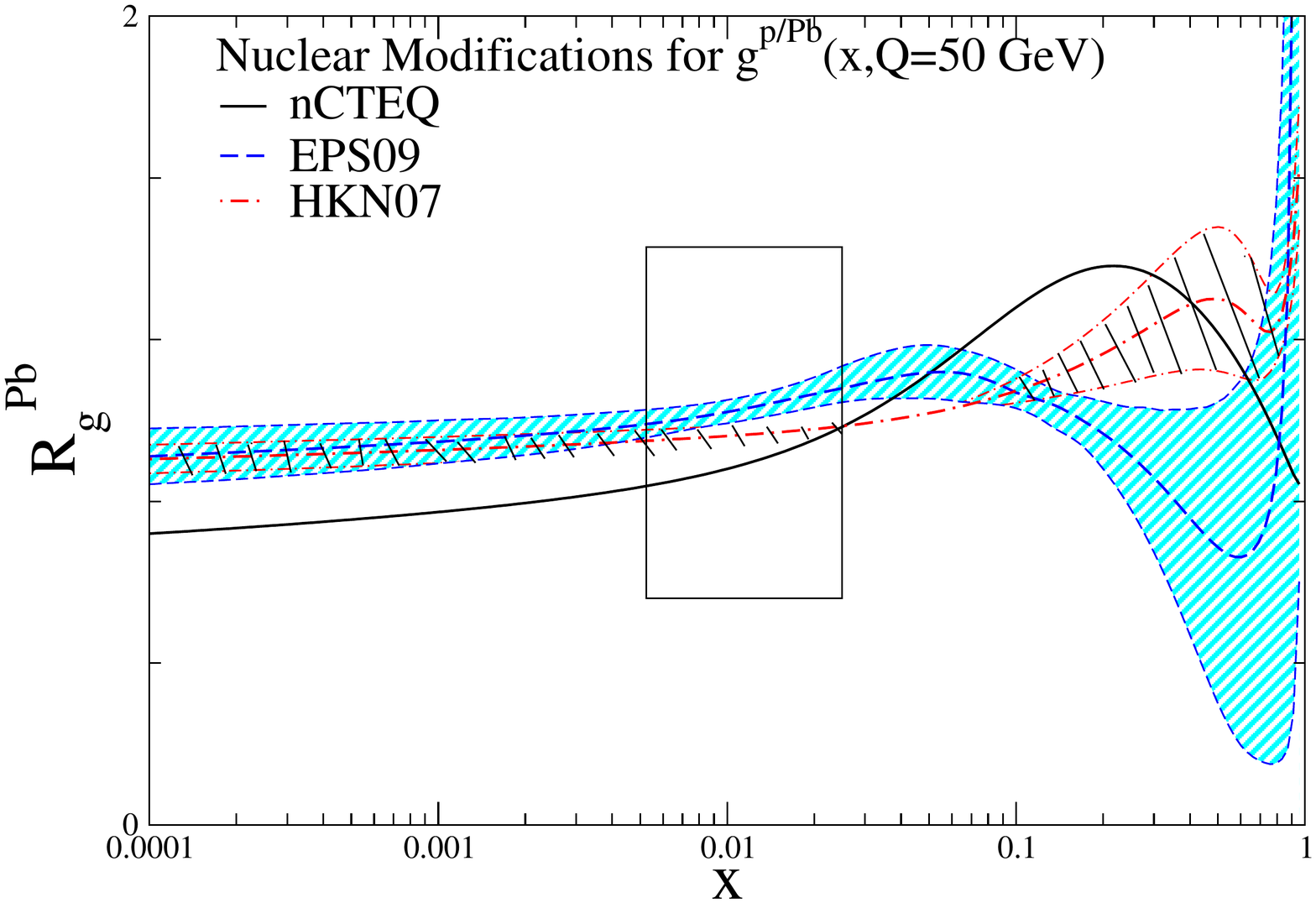}}}
 \put(220,130){\mbox{\includegraphics[angle=270,scale=0.26,trim= 2cm 0cm 0cm 0cm,clip]{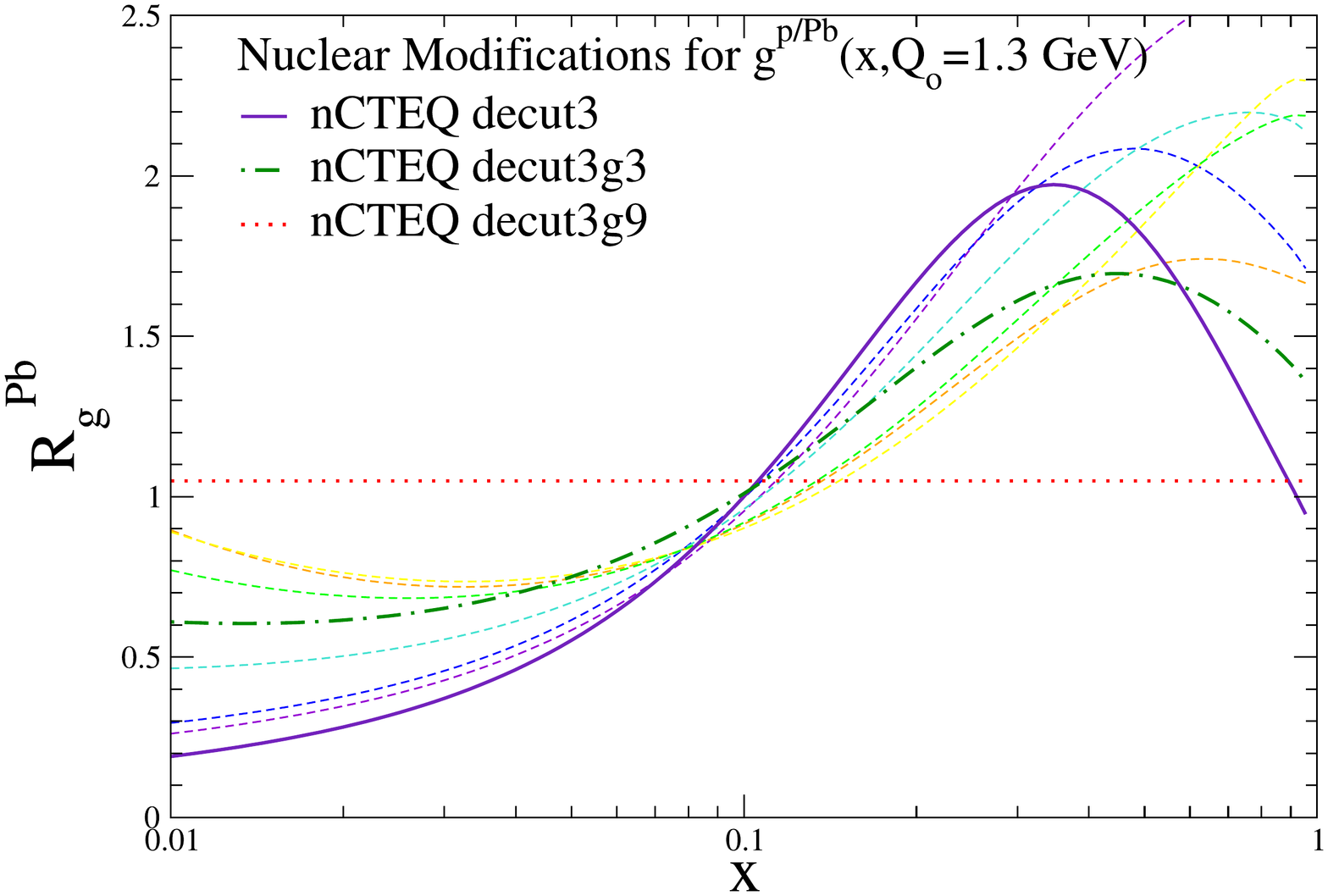}}}
\end{picture}
  \vspace {-1cm}
\caption{a) $R_g^{Pb}(x,Q=x\sqrt{S}/2 \sim p_T)$ for nCTEQ decut3, decut3g3, decut3g9, EPS09 + error band, HKN07 + error band. The box exemplifies the $x$-region probed at the LHC. b) $R_g^{Pb}(x,Q=x\sqrt{S}/2 \sim p_T)$  for different nCTEQ decut3, decut3g1-decut3g9.}
\label{fig:nPDF}
\end{center}
\end{figure*}
%%%
\vspace{-1.3cm}
\begin{figure}[ht]
 \begin{center}
  \includegraphics[angle=0,scale=0.26,trim= 0cm 0cm 0cm 3cm,clip]{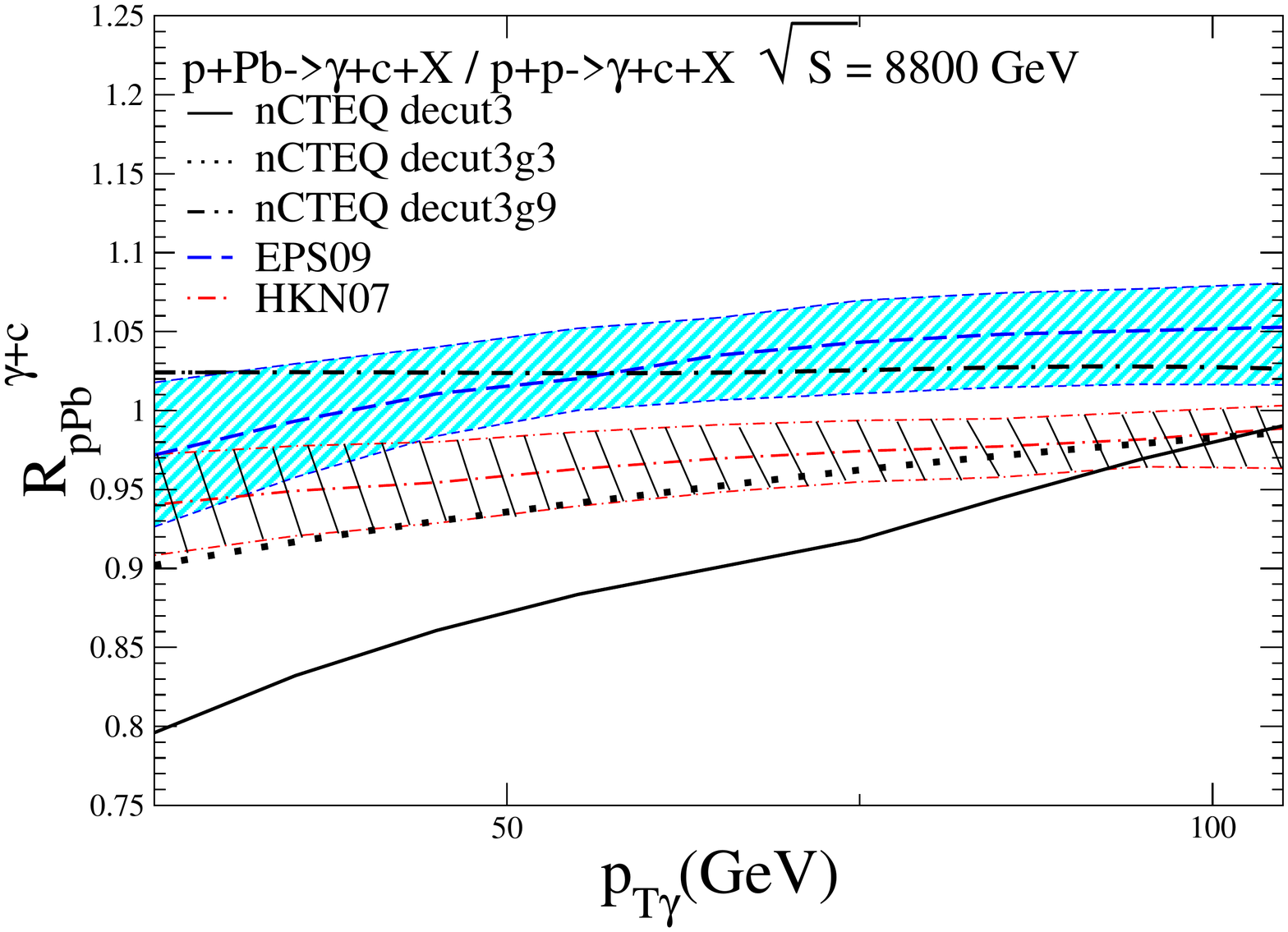} 
  \includegraphics[angle=0,scale=0.26,trim= 0cm 0cm 0cm 3cm,clip]{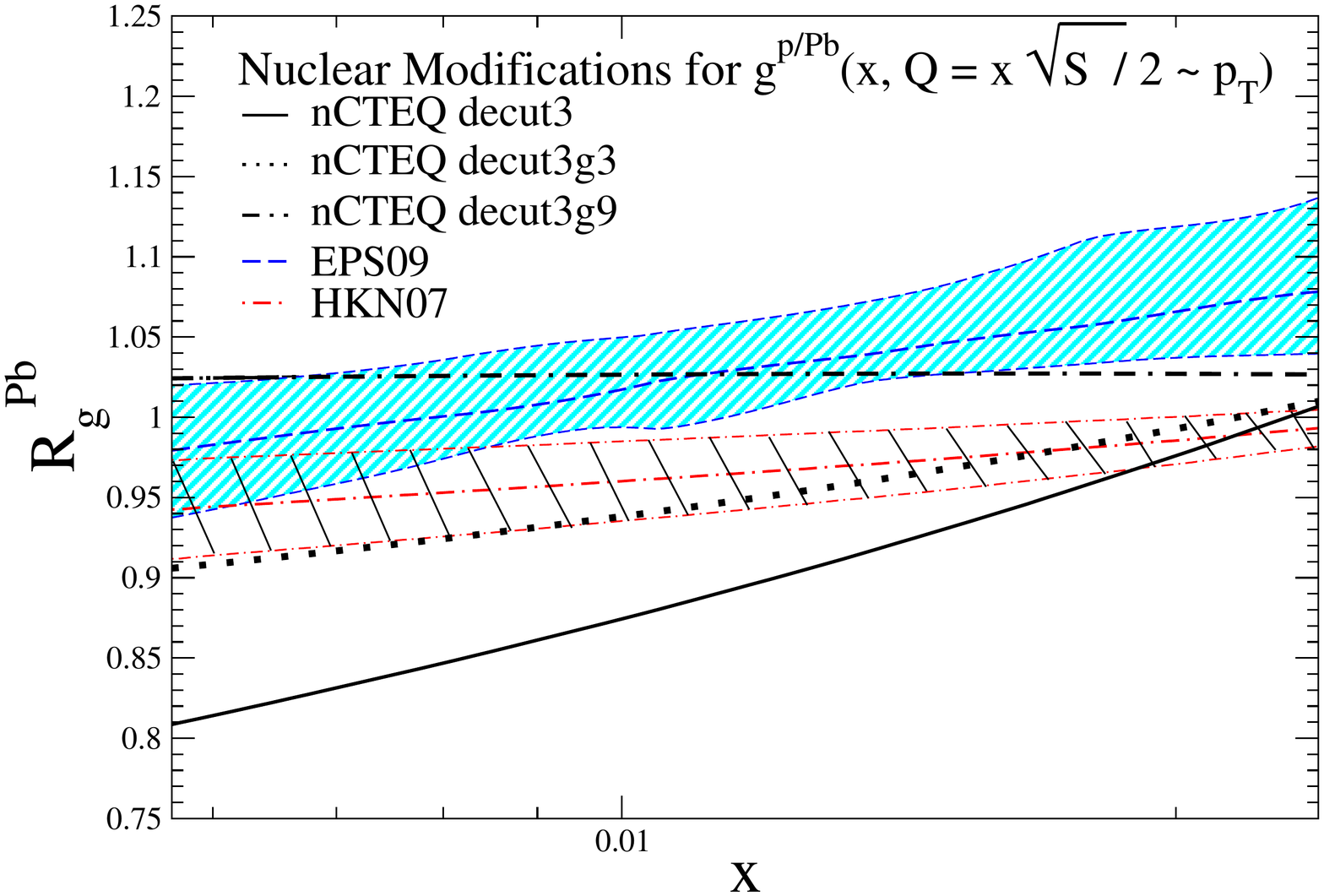} 
  \vspace {-0.78cm}
  \caption{a) $R_{pPb}^{\gamma+c}$ at LHC within ALICE EMCal acceptances, 
using nCTEQ decut3, decut3g3, decut3g9, EPS09 + error band, HKN07 + error band. 
b) $R_g^{Pb}(x,Q=x\sqrt{S}/2 \sim p_T)$ in the $x$ region probed at the LHC.}
  \label{fig:RxsecLHC}
 \end{center}
\end{figure}
\vspace{-0.85cm}
Unlike the PDF for a gluon inside a free proton, the nuclear gluon PDF is largely unconstrained due to the dearth of available data.  Currently, only the NMC structure function data ($F_2^D(x,Q^2)$ and $F_2^{Sn}/F_2^C(x,Q^2)$) impose weak constraints on the gluon nPDF in the $x$-range $0.02\lesssim x\lesssim 0.2$\footnote{The EPS09 fit also includes data on $\pi^0$ production at RHIC.}, so that a precise determination is not possible. 
 This large uncertainty in $g^A(x,Q^2)$ is presented by the nuclear modification factor to the gluon nPDF, $R_g^{Pb}(x,Q)=g^{p/Pb}(x,Q)/g^p(x,Q)$,  in Fig.\ref{fig:nPDF}a)  where a comparison between the different nuclear PDF sets currently available (nCTEQ \cite{Schienbein:2007fs,Schienbein:2009kk,Kovarik:2010uv}, HKN07 \cite{Hirai:2007sx}, 
EPS09 \cite{Eskola:2009uj}) is shown.  Fig.\ref{fig:nPDF}b) furthermore shows different fits with equally good $\chi^2$, whose spread represents a lower limit on the uncertainty associated with the nCTEQ set\footnote{These fits are available at http://projects.hepforge.org/ncteq/ .}.  The need for measurements of processes sensitive to the gluon nPDF is evident.  Here we point out that $\gamma + Q$ production is  an excellent probe of $g^A(x,Q^2)$, and can serve as one such process, as evidenced by Fig.9 and Fig.10 in Ref. \cite{Stavreva:2010mw}.  Fig.9 shows the differential cross-section for both $\gamma+c$ and $\gamma+b$ at $\sqrtsnn=8.8$~TeV  for $p-Pb$ collisions at ALICE EMCal acceptances. The anticipated event rate (before experimental efficiencies)  is sufficiently large for a measurement ($N^{p Pb}_{\gamma+c}=11900$, $N^{p Pb}_{\gamma+b}=2270$).  In Fig.10 the subprocess contributions to $d\sigma^{pPb}_{\gamma+c}/dp_{T\gamma}$ are presented, with $g-Q$ and $g-g$ being the dominant ones; for more details see Ref. \cite{Stavreva:2010mw}.  The sensitivity to the gluon nPDF further shows up in the nuclear modification factor to the cross-section, $R_{pPb}^{\gamma+c}={1\over 208}{d\sigma/dp_{T\gamma}(p{\rm Pb} \rightarrow \gamma+c+X)\over d\sigma/dp_{T\gamma}(pp \rightarrow \gamma+c+X)}$ in Fig.\ref{fig:RxsecLHC}a), when compared to $R_g^{Pb}(x,Q)$ in Fig.\ref{fig:RxsecLHC}b). It can clearly be seen by juxtaposing Fig.\ref{fig:RxsecLHC}a) and Fig.\ref{fig:RxsecLHC}b) that $R_{pPb}^{\gamma+c}$ follows closely $R_g^{Pb}$ in the region of $x$ probed at the LHC for each nPDF set.  Therefore we can conclude that this process is an excellent candidate for constraining the gluon nuclear distribution as a measurement of the prompt photon + heavy-quark jet process with appropriately small error bars  will be able to distinguish between the three different nPDF sets.  
\vspace{-0.6cm}
\section{Heavy Quark Energy Loss in $\gamma +Q$ Production} 
\label{sec:AA} 
\vspace{-0.2cm}
The study of two-particle final states in heavy-ion collisions provides a much more versatile access to quantifying the energy loss in the Quark Gluon Plasma (QGP), as compared to the study of a single inclusive process.  This is further the case if one of the final-state particles is medium insensitive, in particular the study of $\gamma +${\it jet} \cite{Wang:1996yh} or $\gamma+${\it  hadron} \cite{Arleo:2004xj} correlations helps to evaluate the amount of quenching experienced by jets as they traverse the medium while the photon's energy serves as a gauge of the initial parton energy.  Here, the focus on the associated production of $\gamma +${\it heavy-quark jet} can help clarify the energy loss in the heavy quark sector.  Currently, due to the dead cone effect a definite hierarchy of the energy loss is expected, $\epsilon_q > \epsilon_c > \epsilon_b$, with the heavier quarks losing less energy \cite{Dokshitzer:2001zm}.  This hierarchy remains to be clarified experimentally, and prompt photon + heavy-quark jet production is a natural and promising process for this verification.

\begin{figure}[ht]
 \begin{center}
  \includegraphics[angle=0,scale=0.26,trim= 0cm 0cm 0cm 3cm,clip]{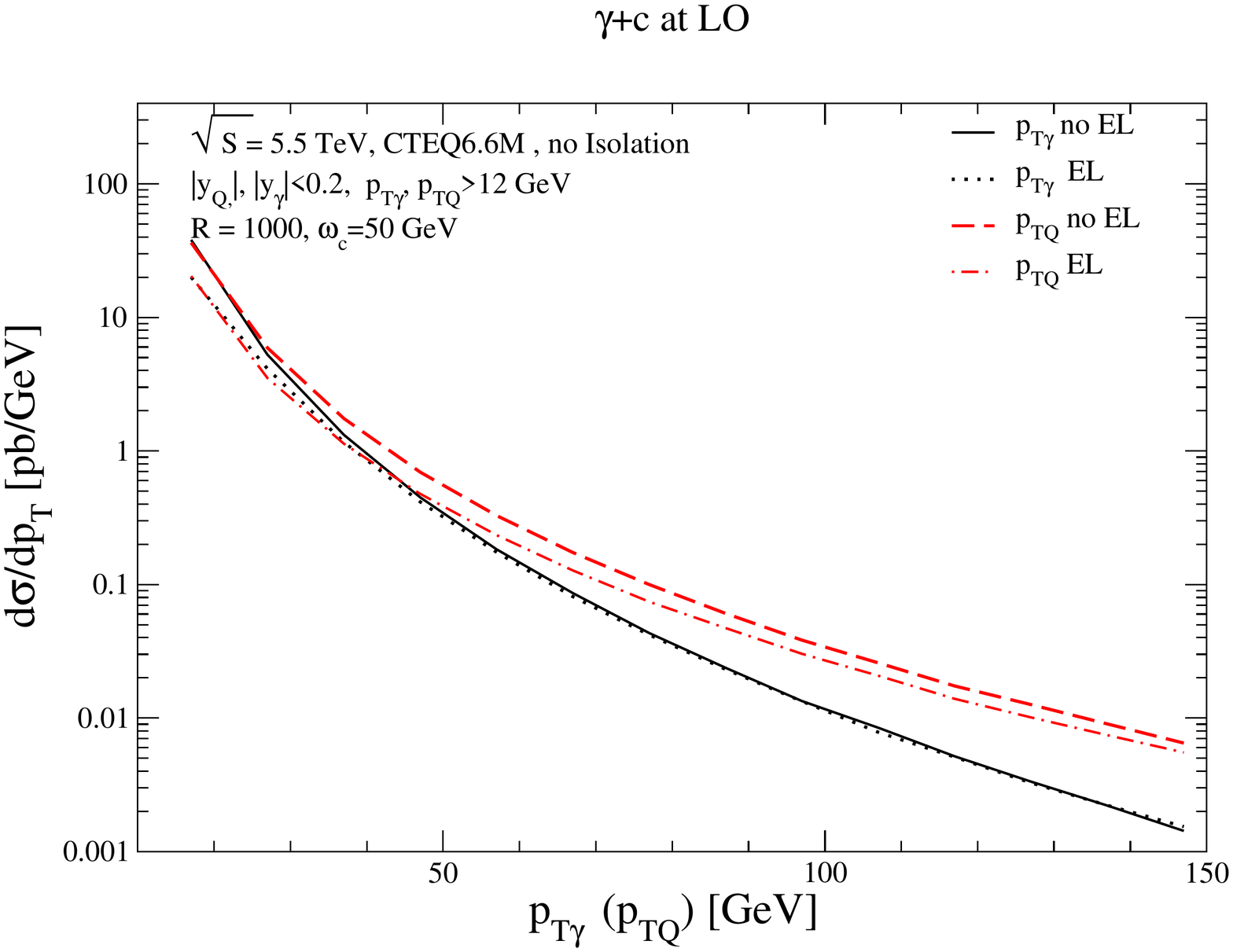} 
  \includegraphics[angle=0,scale=0.26,trim= 0cm 0cm 0cm 3cm,clip]{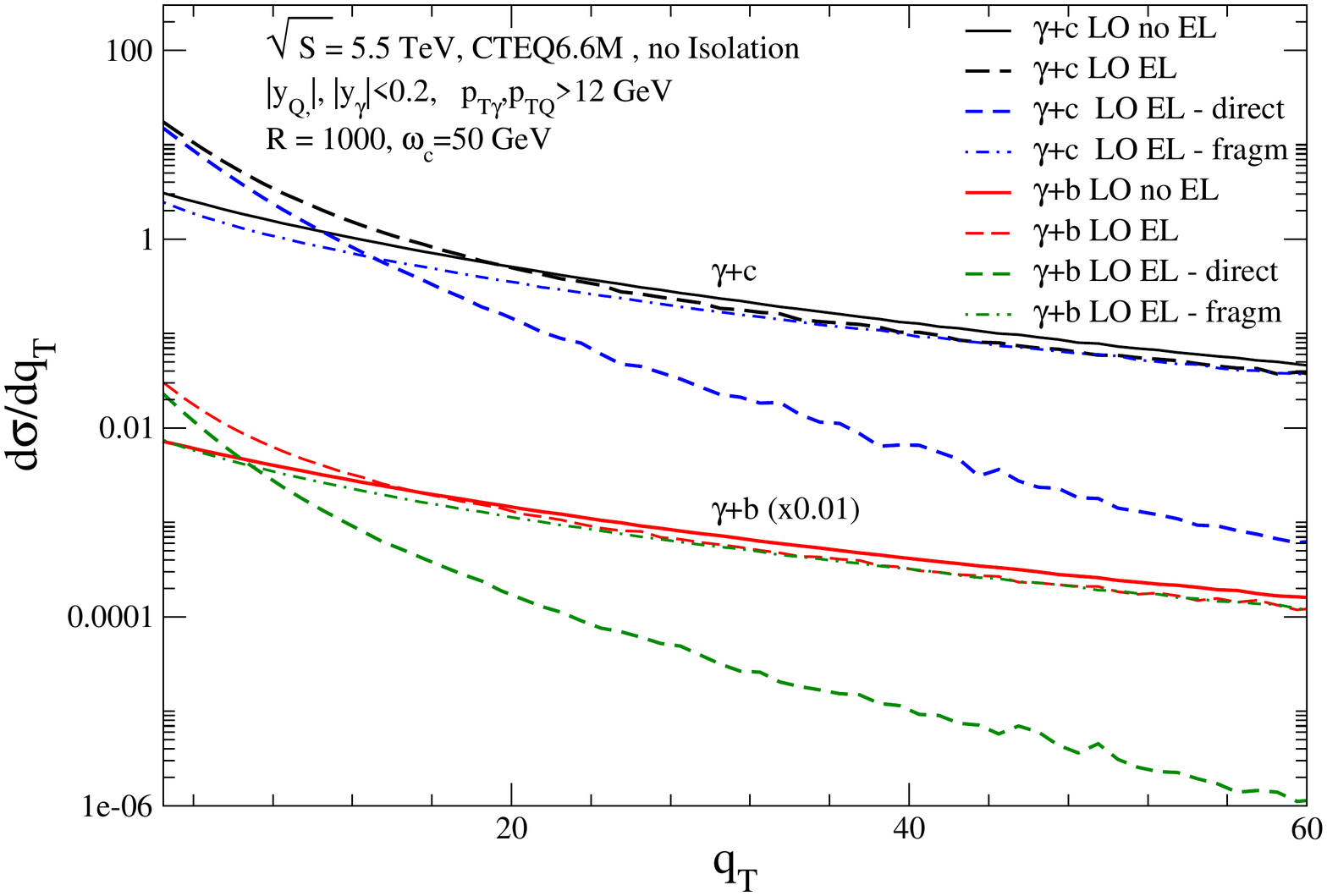} 
  \vspace {-0.8cm}
  \caption{a) The LO $\gamma+c$ differential cross-section versus: $p_{T\gamma}$ in vacuum (solid line), in medium (dotted line); $p_{TQ}$ in vacuum (dashed line), in medium (dashed-dotted line). b) The LO differential cross-section versus $q_\perp$ for $\gamma+c$ and $\gamma+b$, showing the fragmenation and direct contributions in vacuum and in medium.}
  \label{fig:cross-secEL}
 \end{center}
\end{figure}

\begin{figure}[ht]
 \begin{center}
  \includegraphics[angle=0,scale=0.26,trim= 0cm 0cm 0cm 3cm,clip]{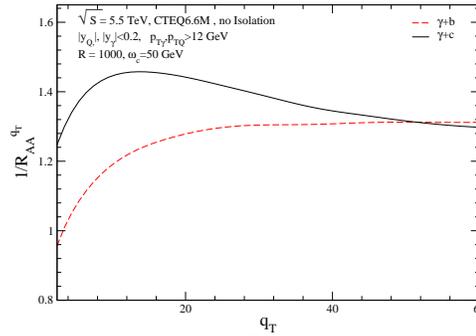} 
  \vspace {-0.9cm}
  \caption{a) The ratio of the LO vacuum fragmentation contribution to the medium fragmentation contribution for $\gamma+c$ (solid line) and $\gamma+b$ (dashed line). }
  \label{fig:RAA}
 \end{center}
\end{figure}

In Fig.\ref{fig:cross-secEL}a) we show the effects of the medium on the leading order (LO) differential cross-section versus $p_{T\gamma}$ and $p_{TQ}$.  The energy loss of the heavy quark, $\epsilon_Q$, is computed on an event by event basis, with the use of the quenching weight obtained perturbatively \cite{Armesto:2005iq}.  The following parameters describing the medium have been used $\hat q =6.25$~GeV/fm$^2$ and $\omega_c=50$~GeV.  The effects show up in the difference between ${d\sigma^{\gamma+c;med}}\over {dp_{TQ}}$ and ${d\sigma^{\gamma+c;vac}}\over {dp_{TQ}}$ \footnote{The small difference between ${{d\sigma^{med}}\over{p_{T\gamma}}}$ and  ${{d\sigma^{vac}}\over{p_{T\gamma}}}$  at low $p_T$ is due to experimental cuts.}.  However, we need not limit ourselves to only one-particle observables as the information obtained by investigating the correlations of the two final state particles  (e.g. photon-jet energy asymmetry, momentum imbalance, photon-jet pair momentum \cite{Arleo:2004xj})  provides a much better handle on the amount of energy loss.  In Fig.\ref{fig:cross-secEL}b) we focus in more detail on the differential cross-section as a function of the photon-jet pair momentum, $q_\perp = | \vec{p}_{T\gamma} + \vec{p}_{TQ} |$.  At LO accuracy, for the direct contribution one has $q_\perp \simeq \epsilon_Q$, whereas for the fragmentation contribution the shift between the $q_T$ spectrum in vacuum versus the one in medium  is given by $<\epsilon_Q>$. Unfortunately when one investigates two-particle observables for this process at LO, only the fragmentation contributions in medium and in vacuum can be compared, as due to the kinematic constraints the direct component in vacuum is non-zero only when $q_T=0$.  In Fig.\ref{fig:RAA} the fragmentation contributions in medium to $\gamma+c$ and $\gamma+b$ normalized to the $p-p$ case are shown. Clearly $\Delta E_c > \Delta E_b$ at small $q_T$, while as  $q_T$ grows the difference disappears, as the quenching weight depends on  $m/E$, which becomes similar for charm and bottom quarks at large $q_T$.  However, definite conclusions can only be drawn after a study at NLO accuracy \cite{wip}. 
\vspace{-0.5cm}
\vspace{-1.3cm}
\vspace{-1.4cm}
\section{Conclusion} 
\vspace{-0.2cm}
Prompt photon + heavy-quark jet production has proven to be an extremely useful and versatile process. It can be employed to constrain the heavy quark PDFs in hadron-hadron collisions, while measurements in  $p-A$ collisions can help constrain the gluon nPDF.  In heavy-ion collisions it can help estimate the quenching experienced by a heavy-quark jet, while also providing access to the mass hierarchy of parton energy loss.

\end{document}